\documentclass[12pt,preprint]{aastex}

\usepackage{color,hyperref}
\definecolor{linkcolor}{rgb}{0,0,0.5}
\hypersetup{colorlinks=true,linkcolor=linkcolor,citecolor=linkcolor,
            filecolor=linkcolor,urlcolor=linkcolor}

\usepackage{url}
\usepackage{algorithmic,algorithm}
\usepackage{amssymb,amsmath}

\newcommand{\arxiv}[1]{\href{http://arxiv.org/abs/#1}{arXiv:#1}}

\usepackage{listings}
\definecolor{lbcolor}{rgb}{0.9,0.9,0.9}
\newcommand{\project}[1]{{\sffamily #1}}
\newcommand{\Python}{\project{Python}}
\newcommand{\numpy}{\project{numpy}}

\newcommand{\github}{\project{GitHub}}
\newcommand{\pip}{\project{pip}}
\newcommand{\acor}{\project{acor}}
\newcommand{\thisplain}{emcee}
\newcommand{\this}{\project{\thisplain}}
\newcommand{\paper}{document}
\newcommand{\license}{MIT License}

\newcommand{\foreign}[1]{\emph{#1}}
\newcommand{\etal}{\foreign{et\,al.}}

\newcommand{\Eq}[1]{Equation~(\ref{eq:#1})}
\newcommand{\eq}[1]{\Eq{#1}}
\newcommand{\eqlabel}[1]{\label{eq:#1}}
\newcommand{\Sect}[1]{Section~\ref{sect:#1}}
\newcommand{\sect}[1]{\Sect{#1}}
\newcommand{\App}[1]{Appendix~\ref{sect:#1}}
\newcommand{\app}[1]{\App{#1}}
\newcommand{\sectlabel}[1]{\label{sect:#1}}
\newcommand{\Algo}[1]{Algorithm~\ref{algo:#1}}
\newcommand{\algo}[1]{\Algo{#1}}
\newcommand{\algolabel}[1]{\label{algo:#1}}

\newcommand{\dd}{\mathrm{d}}

\newcommand{\ensemble}{S}
\newcommand{\colorens}[1]{\ensemble^{(#1)}}
\newcommand{\red}{\colorens{0}}
\newcommand{\blue}{\colorens{1}}
\renewcommand{\vector}[1]{#1}

\newcommand{\pr}[1]{\ensuremath{p(#1)}}
\newcommand{\af}{\ensuremath{a_f}}
\newcommand{\expect}[1]{\left<#1\right>}

\newcommand{\model}{\ensuremath{\vector{\Theta}}}
\newcommand{\data}{\ensuremath{\vector{D}}}
\newcommand{\nuisance}{\ensuremath{\vector{\alpha}}}
\newcommand{\link}{\ensuremath{X}}

\defcitealias{Goodman:2010}{GW10}

\begin{document}

\title{\this: The MCMC Hammer}

\newcommand{\nyu}{2}
\newcommand{\mpia}{3}
\newcommand{\cmu}{4}
\newcommand{\princeton}{5}
\newcommand{\courant}{6}
\author{Daniel~Foreman-Mackey\altaffilmark{1,\nyu},
    David~W.~Hogg\altaffilmark{\nyu,\mpia},
    Dustin~Lang\altaffilmark{\cmu,\princeton},
    Jonathan~Goodman\altaffilmark{\courant}}
\altaffiltext{1}{To whom correspondence should be addressed:
                        \url{danfm@nyu.edu}}
\altaffiltext{\nyu}{Center for Cosmology and Particle Physics,
                        Department of Physics, New York University,
                        4 Washington Place, New York, NY, 10003, USA}
\altaffiltext{\mpia}{Max-Planck-Institut f\"ur Astronomie,
                     K\"onigstuhl 17, D-69117 Heidelberg, Germany}
\altaffiltext{\cmu}{McWilliams Center for Cosmology,
                          Department of Physics, Carnegie Mellon
                          University, 5000 Forbes Ave, Pittsburgh, PA 15213}
\altaffiltext{\princeton}{Princeton University Observatory,
                          Princeton, NJ, 08544, USA}
\altaffiltext{\courant}{Courant Institute, New York University,
                        251 Mercer St., New York, NY 10012, United States}

\begin{abstract}
    We introduce a stable, well tested Python implementation of the affine-%
    invariant ensemble sampler for Markov chain Monte Carlo (MCMC)
    proposed by Goodman \& Weare (2010). The code is open source and has
    already been used in several published projects in the astrophysics
    literature. The algorithm behind \this\ has several advantages over
    traditional MCMC sampling methods and it has excellent performance as
    measured by the autocorrelation time (or function calls per independent sample).
    One major advantage of the algorithm is that it requires hand-tuning of
    only 1 or 2 parameters compared to $\sim N^2$ for
    a traditional algorithm in an $N$-dimensional parameter space. In this
    \paper, we describe the algorithm and the details of our implementation.
    Exploiting the parallelism of the ensemble method,
    \this\ permits \emph{any} user to take advantage of
    multiple CPU cores without extra effort.  The code is available online
    at \url{http://dan.iel.fm/\thisplain} under the \license.
\end{abstract}


~\clearpage

\noindent
\emph{Note: If you want to get started immediately with the \this\ package,
  start at \app{install} on
  page~\pageref{sect:install} or visit the online documentation at
  \url{http://dan.iel.fm/emcee}. If you are sampling with \this\ and having
  low-acceptance-rate or other issues, there is some advice in
  \sect{advice} starting on page~\pageref{sect:advice}.}

\section{Introduction}

Probabilistic data analysis---including Bayesian inference---has
transformed scientific research in the past decade. Many of the most
significant gains have come from numerical methods for approximate
inference, especially Markov chain Monte Carlo (MCMC).  For example,
many problems in cosmology and astrophysics\footnote{The methods and
  discussion in this \paper\ have general applicability, but we will
  mostly present examples from astrophysics and cosmology, the fields
  in which we have most experience} have directly benefited from MCMC
because the models are often expensive to compute, there are many free
parameters, and the observations are usually low in signal-to-noise.

Probabilistic data analysis procedures involve computing and using
either the posterior probability density function (PDF) for the
parameters of the model or the likelihood function. In some cases it
is sufficient to find the maximum of one of these, but it is often
necessary to understand the posterior PDF in detail.  MCMC methods are
designed to sample from---and thereby provide sampling approximations
to---the posterior PDF efficiently even in parameter spaces with large
numbers of dimensions. This has proven useful in too many research
applications to list here but the results from the NASA Wilkinson
Microwave Anisotropy Probe (WMAP) cosmology mission provide a dramatic
example \citep[for example,][]{Dunkley:2005}.

Arguably the most important advantage of Bayesian data analysis is
that it is possible to \emph{marginalize} over nuisance parameters. A
nuisance parameter is one that is required in order to model the
process that generates the data, but is otherwise of little interest.
Marginalization is the process of integrating over all possible values of
the  parameter and hence propagating the effects of uncertainty about
its value into the final result.  Often we wish to marginalize over all
nuisance parameters in a model.  The exact result of marginalization
is the marginalized probability function \pr{\model | \data}
of the set (list or vector) of model parameters
\model\ given the set of observations \data
\begin{equation}
    \eqlabel{marginalization}
    \pr{\model | \data} = \int
        \pr{ \model, \nuisance | \data} \,
        \dd  \nuisance \quad,
\end{equation}
where \nuisance\ is the set (list or vector) of nuisance
parameters. Because the nuisance parameter set \nuisance\ can be very large, this
integral is often extremely daunting.  However, a
MCMC-generated sampling of values $(\model_t,\nuisance_t)$ of the
model and nuisance parameters from the joint distribution $\pr{\model,
  \nuisance | \data}$ automatically provides a sampling of values
$\model_t$ from the marginalized PDF $\pr{\model | \data}$.

In addition to the problem of marginalization, in many problems of
interest the likelihood or the prior is the result of an expensive
simulation or computation. In this regime, MCMC sampling is very
valuable, but it is even \emph{more} valuable if the MCMC algorithm is
efficient, in the sense that it does not require many function
evaluations to generate a statistically independent sample from the
posterior PDF.  The methods presented here are designed for efficiency.

Most uses of MCMC in the astrophysics literature are based on slight
modifications to the Metropolis-Hastings (M--H) method (introduced
below in \sect{algo}).  Each step in a M--H chain is proposed using a
compact proposal distribution centered on the current position of the
chain (normally a multivariate Gaussian or something similar). Since
each term in the covariance matrix of this proposal distribution is an
unspecified parameter, this method has $N\,[N+1]/2$ tuning parameters
(where $N$ is the dimension of the parameter space).  To make matters
worse, the performance of this sampler is very sensitive to these
tuning parameters and there is no fool-proof method for choosing the
values correctly. As a result, many heuristic methods have been
developed to attempt to determine the optimal parameters in a
data-driven way \citep[for
  example,][]{Gregory:2005,Dunkley:2005,Widrow:2008}. Unfortunately,
these methods all require a lengthy ``burn-in'' phase where shorter
Markov chains are sampled and the results are used to tune the
hyperparameters. This extra cost is unacceptable when the likelihood
calls are computationally expensive.

The problem with traditional sampling methods can be visualized by looking
at the simple but highly anisotropic density
\begin{equation}
    \eqlabel{anisotropic}
    p(\mathbf{x}) \propto f \left (-\frac{(x_1-x_2)^2}{2\,\epsilon}
                                        - \frac{(x_1+x_2)^2}{2} \right )
\end{equation}
which would be considered difficult (in the small-$\epsilon$ regime) for
standard MCMC algorithms. In principle, it is possible to tune the
hyperparameters of a M--H sampler to make this sampling converge quickly,
but if the dimension is large and calculating the density
is computationally expensive the tuning procedure becomes intractable.
Also, since the number of parameters scales as $\sim N^2$, this problem gets
much worse in higher dimensions.
\Eq{anisotropic} can, however, be transformed into the much easier problem of
sampling an isotropic density by an \emph{affine transformation} of the form
\begin{equation}
    y_1 = \frac{x_1-x_2}{\sqrt{\epsilon}} \, ,
        \hspace{1cm} y_2 = x_1 + x_2 \quad .
\end{equation}
This motivates affine invariance: an algorithm that is \emph{affine invariant}
performs equally well under all linear transformations; it will therefore be
insensitive to
covariances among parameters.

Extending earlier work by \citet{Christen:2007},
\citet[][hereafter \citetalias{Goodman:2010}]{Goodman:2010} proposed an
affine invariant sampling
algorithm (\sect{algo}) with only two hyperparameters to be tuned for
performance. \citet{Hou:2011} were the first group to implement this
algorithm in astrophysics. The implementation presented here is
an independent effort that has already proved effective in several projects
\citep{sanders2013,reis2013,weisz2013,cieza2013,akeret2012,huppenkothen2012,
monnier2012,morton2012,crossfield2012,roskar2012,bovy2012b,brown2012,
brammer2012,bussmann2012,bovy2012a,lang2012,bovy2012,olofsson2012,dorman2012}.
In what follows, we summarize the
algorithm from \citetalias{Goodman:2010} and the implementation
decisions made in \this. We also describe the small changes
that must be made to the algorithm to parallelize it. Finally, in the
Appendices, we outline the installation, usage and troubleshooting of
the package.

\section{The Algorithm}\sectlabel{algo}

A complete discussion of MCMC methods is beyond the scope of this \paper.
Instead, the interested reader is directed to a classic reference like
\citet{MacKay:2003} and we will summarize some key concepts below.

The general goal of MCMC algorithms is to draw $M$ samples
$\{ \model_i \}$ from
the posterior probability density
\begin{equation}
    \pr{\model, \nuisance | \data} = \frac{1}{Z}\,\pr{\model, \nuisance}
            \, \pr{\data | \model, \nuisance} \quad,
\end{equation}
where the prior distribution $\pr{\model, \nuisance}$ and the likelihood
function $\pr{\data|\model,\nuisance}$ can be relatively easily (but not
necessarily quickly) computed for any particular value of
$(\model_i, \nuisance_i)$.  The normalization $Z=\pr{\data}$ is
independent of $\model$ and $\nuisance$ once we have chosen the form of the
generative model. This means that it is possible
to sample from \pr{\model, \nuisance | \data} without computing $Z$ ---
unless one would like to compare the validity of two different generative
models. This is important because $Z$ is generally very expensive to
compute.

Once the samples
produced by MCMC are available, the marginalized constraints on $\model$
can be approximated by
the histogram of the samples projected into the parameter subspace spanned
by $\model$. In particular, this implies that the
expectation value of a function of the model parameters $f(\model)$ is
\begin{equation}
    \expect{f(\model)} = \int
    \pr{\model|\data}
    \, f(\model) \, \dd\model
    \,\approx\, \frac{1}{M} \sum_{i=1} ^M f(\model_i) \quad.
\end{equation}
Generating the samples $\model_i$ is a non-trivial process unless
$\pr{\model, \nuisance, \data}$ is a very specific analytic distribution
(for example, a Gaussian). MCMC is a procedure for generating a random walk
in the parameter space that, over time, draws a representative set
of samples from the distribution. Each point in a Markov chain
$\link (t_i) = [\model_i, \nuisance_i]$
depends only on the position of the previous step $\link (t_{i-1})$.

\paragraph{The Metropolis-Hastings (M--H) Algorithm}

The simplest and most commonly used MCMC algorithm is the M--H method
\citep[\algo{mh};][]{MacKay:2003,Gregory:2005,Press:2007,Hogg:2010}.
The iterative procedure is as follows: (1) given a position
$X(t)$ sample a proposal position $Y$ from the transition distribution
$Q(Y; X(t))$, (2) accept this proposal with probability
\begin{equation}
    \mathrm{min} \left( 1,\,
            \frac{\pr{\vector{Y} | \data}}{\pr{\vector{X}(t) | \data}} \,
            \frac{Q(X(t); Y)}{ Q(Y;X(t))}  \right) \quad.
\end{equation}
The transition distribution $Q(Y; X(t))$ is an
easy-to-sample probability distribution for the proposal $Y$ given
a position $X(t)$.
A common parameterization of $Q(Y; X(t))$ is a multivariate Gaussian
distribution centered on $X(t)$ with a general covariance tensor that has
been tuned for performance.
It is worth emphasizing that if this step is accepted $X(t+1) = Y$; Otherwise,
the new position is set to the previous one $X(t+1) = X(t)$ (in other
words, the position $X(t)$ is \emph{repeated in the chain}).

The M--H algorithm converges (as $t \to \infty$) to a stationary set of
samples from the distribution but there are many algorithms with faster
convergence and varying levels of implementation difficulty.
Faster convergence is preferred because of the reduction of computational
cost due to the smaller number of likelihood computations necessary to obtain
the equivalent level of accuracy. The inverse convergence rate can be
measured by the autocorrelation function and more specifically, the integrated
autocorrelation time (see \sect{tests}). This quantity is an estimate of the
number of steps needed in the chain in order to draw independent samples from
the target density. A more efficient chain has a shorter
autocorrelation time.

\begin{algorithm}
\caption{The procedure for a single Metropolis-Hastings MCMC step.
    \algolabel{mh}}
\begin{algorithmic}[1]

\STATE Draw a proposal $Y \sim Q (Y; X(t))$
\STATE $q \gets [\pr{\vector{Y}} \, Q(X(t); Y)]
        / [\pr{\vector{X}(t)} \, Q(Y;X(t))]$%
            \hspace{1cm}{\footnotesize\it // This line is generally expensive}
\STATE $r \gets R \sim [0, 1]$
\IF{$r \le q$}
    \STATE $\vector{X}(t+1) \gets \vector{Y}$
\ELSE
    \STATE $\vector{X}(t+1) \gets \vector{X}(t)$
\ENDIF

\end{algorithmic}
\end{algorithm}

\paragraph{The stretch move}

\citetalias{Goodman:2010} proposed an affine-invariant ensemble sampling
algorithm informally called the ``stretch move.'' This algorithm
significantly outperforms standard M--H methods producing independent
samples with a much shorter autocorrelation time (see \sect{acor} for
a discussion of the autocorrelation time). For completeness and for
clarity of notation, we summarize the algorithm here and refer the interested
reader to the original paper for more details. This method involves
simultaneously evolving an ensemble of $K$ \emph{walkers}
$\ensemble = \{ \vector{X_k} \}$ where the proposal
distribution for one walker $k$ is based on the current positions of the
$K-1$ walkers in the \emph{complementary ensemble}
$\ensemble_{[k]} = \{ \vector{X_j}, \, \forall j \ne k \}$. Here, ``position''
refers to a vector in the $N$-dimensional, real-valued parameter space.

To update the position of a walker at position $\vector{X_k}$,
a walker $X_j$ is drawn randomly from the remaining walkers $\ensemble_{[k]}$
and a new position is proposed:
\begin{equation}
    \eqlabel{proposal}
    \vector{X_k} (t) \to \vector{Y} = \vector{X_j}
            + Z \, [\vector{X_k} (t) - \vector{X_j}]
\end{equation}
where $Z$ is a random variable drawn from a distribution $g(Z = z)$.
It is clear that if $g$ satisfies
\begin{equation}
    g(z^{-1}) = z \, g(z),
\end{equation}
the proposal of \eq{proposal} is symmetric. In this case, the chain will
satisfy detailed balance if the proposal is accepted with probability
\begin{equation}
    \eqlabel{acceptance}
    q = \min \left( 1,\, Z^{N-1} \,
                \frac{\pr{\vector{Y}}}{\pr{\vector{X_k} (t)}} \right) \quad,
\end{equation}
where $N$ is the dimension of the parameter space. This procedure is then
repeated for each walker in the ensemble \emph{in series} following the
procedure shown in \algo{goodman}.

\citetalias{Goodman:2010} advocate a particular form of $g(z)$, namely
\begin{equation}
    \eqlabel{goodmanprop}
    g(z) \propto \left \{ \begin{array}{ll}
        \displaystyle\frac{1}{\sqrt{z}} & \mathrm{if}\, z\in
                        \left [ \displaystyle\frac{1}{a}, a \right ], \\
        0 & \mathrm{otherwise}
    \end{array} \right .
\end{equation}
where $a$ is an adjustable scale parameter that \citetalias{Goodman:2010} set
to 2.

\begin{algorithm}
\caption{A single stretch move update step from \citetalias{Goodman:2010}
    \algolabel{goodman}}
\begin{algorithmic}[1]
\FOR{$k = 1, \ldots, K$}
    \STATE Draw a walker $X_j$ at random from the complementary ensemble %
        $\ensemble_{[k]}(t)$
    \STATE $z \gets Z \sim g(z)$, \Eq{goodmanprop}
    \STATE $\vector{Y} \gets \vector{X_j} %
                + z \, [ \vector{X_k} (t) - \vector{X_j}]$
    \STATE $q \gets z^{N-1} \, p(Y)/p(X_k(t))$ \label{line:hard}%
        \hspace{1cm}{\footnotesize\it // This line is generally expensive}
    \STATE $r \gets R \sim [0, 1]$
    \IF{$r \le q$, \eq{acceptance}}
        \STATE $X_k(t+1) \gets Y$
    \ELSE
        \STATE $X_k(t+1) \gets X_k(t)$
    \ENDIF
\ENDFOR
\end{algorithmic}
\end{algorithm}

\paragraph{The parallel stretch move}

It is tempting to parallelize the stretch move algorithm by
simultaneously advancing each walker based on the state of the ensemble
instead of evolving the walkers in series. Unfortunately, this subtly
violates detailed balance. Instead, we must split the full ensemble
into two subsets
($\red = \{ \vector{X_k}, \, \forall k = 1, \ldots, K/2 \}$ and
$\blue = \{ \vector{X_k}, \, \forall k = K/2+1, \ldots, K \}$) and
simultaneously update all the walkers in $\red$
--- using the stretch move procedure from \algo{goodman} ---
based \emph{only} on the positions of the walkers in the other set
($\blue$). Then, using the new positions $\red$,
we can update $\blue$. In this case, the outcome is a valid step
for all of the walkers. The pseudocode for
this procedure is shown in \algo{parallel}. This code is similar to
\algo{goodman} but now the computationally expensive inner loop
(starting at line~\ref{line:parallelloop} in \algo{parallel}) can be run in
parallel.

The performance of this method --- quantified by the autocorrelation time ---
is comparable to the serial stretch move algorithm but the fact that one
can now take advantage of generic parallelization makes it
extremely powerful.

\begin{algorithm}
\caption{The parallel stretch move update step
    \algolabel{parallel}}
\begin{algorithmic}[1]
\FOR{$i \in \{0, 1\}$}
    \FOR{$k = 1, \ldots, K/2$} \label{line:parallelloop}
        \STATE {\footnotesize\it // This loop can now be done in parallel %
            for all $k$}
        \STATE Draw a walker $\vector{X_j}$ at random from the complementary %
            ensemble $\colorens{\sim i} (t)$
        \STATE $\vector{X_k} \gets \colorens{i}_k$
        \STATE $z \gets Z \sim g(z)$, \Eq{goodmanprop}
        \STATE $\vector{Y} \gets \vector{X_j}
                + z \, [ \vector{X_k} (t) - \vector{X_j}]$
        \STATE $q \gets z^{n-1} \, p(\vector{Y})/p(\vector{X}_k(t))$
        \STATE $r \gets R \sim [0, 1]$
        \IF{$r \le q$, \eq{acceptance}}
            \STATE $\vector{X_k} (t+\frac{1}{2}) \gets \vector{Y}$
        \ELSE
            \STATE $\vector{X_k} (t+\frac{1}{2}) \gets \vector{X_k}(t)$
        \ENDIF
    \ENDFOR
    \STATE $t \gets t+\frac{1}{2}$
\ENDFOR

\end{algorithmic}
\end{algorithm}

\section{Tests} \sectlabel{tests}

Judging the convergence and performance of an algorithm is a
non-trival problem and there is a huge associated literature
\citep[see, for example,][for a review]{Cowles:1996}. In astrophysics,
spectral methods have been used extensively \citep[for
example][]{Dunkley:2005}. Below, we advocate for one such method: the
autocorrelation time. The autocorrelation time is especially applicable
because it is an affine invariant measure of the performance.

First,
however, we should take note of another extremely important measurement:
the acceptance fraction \af. This is the fraction of proposed steps that are
accepted. There appears to be no agreement on the optimal acceptance rate
but it is clear that both extrema are unacceptable. If $\af \sim 0$, then
nearly all proposed steps are rejected, so
the chain will have very few independent samples and the sampling will not be
representative of the target density. Conversely, if $\af \sim 1$ then nearly
all steps are accepted and the
chain is performing a random walk with no regard for the target density so
this will also not produce representative samples. As a rule of thumb, the
acceptance fraction should be between $0.2$ and $0.5$
\citep[for example,][]{Gelman:1996}. For the M--H algorithm,
these effects can generally be counterbalanced by decreasing (or increasing,
respectively) the eigenvalues of the proposal distribution covariance. For
the stretch move, the parameter $a$ effectively controls the step size so
it can be used to similar effect. In our tests, it has never been
necessary to use a value of $a$ other than $2$, but we make no guarantee that
this is the optimal value.

\paragraph{Autocorrelation time} \sectlabel{acor}

The autocorrelation time is a direct measure of the number of evaluations of
the posterior PDF required to produce independent samples of the target
density. \citetalias{Goodman:2010} show that the stretch-move algorithm
has a significantly shorter autocorrelation time on several non-trivial
densities. This means that fewer PDF computations are required---compared
to a M--H sampler---to produce the same number of independent samples.

The autocovariance function of a time series $\vector{X} (t)$ is
\begin{equation}
    C_f (T) = \lim_{t \to \infty} \mathrm{cov}
        \left [ f\left (\vector{X}(t+T) \right ),
            f\left (\vector{X}(t) \right ) \right ].
\end{equation}
This measures the covariances between samples at a time lag $T$. The
value of $T$ where $C_f(T) \to 0$ measures the number of samples that
must be taken in order to ensure independence. In particular, the
relevant measure of sampler efficiency is the integrated autocorrelation
time
\begin{equation}
    \tau_f = \sum_{T=-\infty} ^{\infty} \frac{C_f(T)}{C_f(0)}
        = 1+2\sum_{T=1} ^{\infty} \frac{C_f(T)}{C_f(0)}.
\end{equation}
In practice, one can estimate $C_f (T)$ for a Markov chain of $M$ samples as
\begin{equation}
    C_f (T) \approx \frac{1}{M-T} \sum_{m=1}^{M-T}
        \left [ f(X(T+m)) - \expect{f} \right ] \,
        \left [ f(X(m)) - \expect{f} \right ].
\end{equation}

We advocate for the autocorrelation time as a measure of sampler
performance for two main reasons. First, it measures a quantity
that \emph{we are actually interested in} when sampling in practice.
The longer the autocorrelation time, the more samples that we must
generate to produce a representative sampling of the target
density. Second, the autocorrelation time is affine invariant. Therefore,
it is reasonable to measure the performance and diagnose the convergence
of the sampler on densities with different levels of anisotropy.

\this\ can optionally calculate the autocorrelation time using the Python
module \project{acor}\footnote{\url{http://github.com/dfm/acor}} to estimate
the autocorrelation time. This module is a direct port of the original
algorithm \citepalias[described by][]{Goodman:2010} and implemented by those
authors in
C++.\footnote{\url{http://www.math.nyu.edu/faculty/goodman/software/acor}}

\section{Discussion \& Tips}\sectlabel{advice}

The goal of this project has been to make a sampler that is a useful
tool for a large class of data analysis problems---a ``hammer'' if you
will.  If development of statistical and data-analysis understanding
is the key goal, a user who is new to MCMC benefits enormously by
writing her or his own Metropolis--Hastings code (\algo{mh}) from
scratch before downloading \this.  For typical problems, the
\this\ package will perform better than any home-built M--H code (for
all the reasons given above), but the intuitions developed by writing
and tuning a self-built MCMC code cannot be replaced by reading this
document and running this pre-built package.  That said, once those
intuitions are developed, it makes sense to switch to \this\ or a
similarly well engineered piece of code for performance on large
problems.

Ensemble samplers like \this\ require some thought for initialization.
One general approach is to start the walkers at a sampling of the
prior or spread out over a reasonable range in parameter space.
Another general approach is to start the walkers in a very tight
$N$-dimensional ball in parameter space around one point that is
expected to be close to the maximum probability point.  The first is
more objective but, in practice, we find that the latter is much more
effective if there is any chance of walkers getting stuck in low
probability modes of a multi-modal probability landscape.  The walkers
initialized in the small ball will expand out to fill the relevant
parts of parameter space in just a few autocorrelation times.  A third
approach would be to start from a sampling of the prior, and go
through a ``burn-in'' phase in which the prior is transformed
continuously into the posterior by increasing the ``temperature.''
Discussion of this kind of annealing is beyond the scope of this
document.

It is our present view that autocorrelation time is the best indicator
of MCMC performance (the shorter the better), but there are several
proxies.  The easiest and simplest indicator that things are going
well is the acceptance fraction; it should be in the 0.2 to 0.5 range
\citep[there are theorems about this for specific problems;
for example][]{Gelman:1996}.  In principle,
if the acceptance fraction is too low, you can raise it by decreasing
the $a$ parameter; and if it is too high, you can reduce it by
increasing the $a$ parameter.  However, in practice, we find that
$a=2$ is good in essentially all situations.  That means that when
using \this\ \emph{if the acceptance fraction is getting very low,
  something is going very wrong}.  Typically a low acceptance fraction
means that the posterior probability is multi-modal, with the modes
separated by wide, low probability ``valleys.''  In situations like
these, the best idea (though expensive of human time) is to split the
space into disjoint single-mode regions and sample each one
independently, combining the independently sampled regions
``properly'' (also expensive, and beyond the scope of this document)
at the end.  In previous work, we have advocated clustering methods to
remove multiple modes \citep{Hou:2011}.  These work well when the
different modes have \emph{very} different posterior probabilities.

Another proxy for short autocorrelation time is large expected or mean
squared jump distance (ESJD; \citealt{Pasarica:2010}).  The higher the
ESJD the better; if walkers move (in the mean) a large distance per
chain step then the
autocorrelation time will tend to be shorter.  The ESJD is not an
affine-invariant measure of performance, and it doesn't have a
trivial interpretation in terms of independent samples, so we prefer
the autocorrelation time in principle.  In practice, however, because
the ESJD is a simple expectation value it can be more robustly
evaluated on short chains.

With \this\  you want (in general) to \emph{run with a
  large number of walkers}, like hundreds.  In principle, there is no
reason not to go large when it comes to walker number, until you hit
performance issues.  Although each step takes twice as much compute
time if you double the number of walkers, it also returns to you twice
as many independent samples per autocorrelation time.  So go large.
In particular, we have found that---in almost all cases of low
acceptance fraction---increasing the number of walkers improves the
acceptance fraction.  The one disadvantage of having large numbers of
walkers is that the burn-in phase (from initial conditions to
reasonable sampling) can be slow; burn-in time is a few
autocorrelation times; the total run time for burn-in scales with the
number of walkers.  These considerations, all taken together, suggest
using the smallest number of walkers for which the acceptance fraction
during burn-in is good, or the number of samples you want out at the
end (see below), whichever is \emph{greater}.  A more ambitious
project would be to increase the number of walkers after burn-in; this
requires thought beyond the scope of this document; it can be
accomplished by burning in a set of small ensembles and then merging
them into a big ensemble for the final run.

One mistake many users of MCMC methods make is to take \emph{too many}
samples!  If all you want your MCMC to do is produce one- or
two-dimensional error bars on two or three parameters, then you only
need dozens of independent samples.  With ensemble sampling, you
get this from a \emph{single snapshot} or single timestep, provided
that you are using dozens of walkers (and we would recommend that you
use hundreds in most applications).  The key point is that \emph{you
  should run the sampler for a few (say 10) autocorrelation times.}
Once you have run that long, no matter how you initialized the
walkers, the set of walkers you obtain at the end should be an
independent set of samples from the distribution, of which you rarely
need many.

Another common mistake, of course, is to run the sampler for \emph{too
  few} steps.  You can identify that you haven't run for enough steps
in a couple of ways: If you plot the parameter values in the ensemble
as a function of step number, you will see large-scale variations over
the full run length if you have gone less than an autocorrelation
time.  You will also see that if you try to measure the
autocorrelation time (with, say, \acor), it will give you a time that
is always a significant fraction of your run time; it is only when the
correlation time is much shorter (say by a factor of 10) than your run
time that you are sure to have run long enough.  The danger of both of
these methods---an unavoidable danger at present---is that you can
have a huge dynamic range in contributions to the autocorrelation
time; you might think it is 30 when in fact it is 30\,000, but you
don't ``see'' the 30\,000 in a run that is only 300 steps long.  There
is not much you can do about this; it is generic when the posterior is
multi-modal: The autocorrelation time within each mode can be short but
the mode--mode migration time can be long.  See above on low
acceptance ratio; in general when your acceptance ratio gets low your
autocorrelation time is very, very long.

There are some cases where \this\ won't perform as well as some
more specialized sampling techniques. In particular, when the target density
is multi-modal, walkers can become ``stuck'' in different modes. When
this happens, the vector between walkers is no longer a good proposal
direction. In these cases, the acceptance fraction and autocorrelation
time can deteriorate quickly. While this is a fairly general problem, we
find that in many applications the effect isn't actually very important.
That being said, there are some problems where higher-end machinery
\citep[such as][Hou et al.\ forthcoming]{dnest} is necessary \citep[see, for
example,][]{brewer2012,vh2013}.

Another limitation to the stretch move and moves like it is
that they implicitly assume that the parameters can be assembled into
a vector-like object on which linear operations can be performed.
This is not (trivially) true for parameters that have non-trivial
constraints, like parameters that must be integer-valued or
equivalent, or parameters that are subject to deterministic non-linear
constraints. Sometimes these issues can be avoided by
reparameterization, but in some cases, samplers like \this\ will not
be useful, or might require clever or interesting improvements. The
\this\ package is open-source software; please push us changes!

\acknowledgments It is a pleasure to thank
Eric Agol (UWash),
Jo Bovy (IAS),
Brendon Brewer (Auckland),
Jacqueline Chen (MIT),
Alex Conley (Colorado),
Will Meierjurgen Farr (Northwestern),
Andrew Gelman (Columbia),
John Gizis (Delaware),
Fengji Hou (NYU),
Jennifer Piscionere (Vanderbilt),
Adrian Price-Whelan (Columbia),
Hans-Walter Rix (MPIA),
Jeremy Sanders (Cambridge),
Larry Widrow (Queen's), and
Joe Zuntz (Oxford)
for helpful contributions to the ideas and code presented here.
This project was partially supported by the NSF (grant AST-0908357), NASA
(grant NNX08AJ48G), and DOE (grant DE-FG02-88ER25053).
\this\ makes use of the open-source Python \numpy\ package.

\clearpage
\appendix
\section{Installation}\sectlabel{install}

The easiest way to install \this\ is using
\pip\footnote{\url{http://pypi.python.org/pypi/pip}}. Running the command
\begin{lstlisting}
% pip install emcee
\end{lstlisting}
at the command line of a UNIX-based system will install the package in your
\Python\ path.
If you would like to install for all users, you might
need to run the above command with superuser permissions.
In order to use \this, you must also have
\numpy\footnote{\url{http://numpy.scipy.org}} installed (this can also be
achieved using \pip\ on most systems).
\this\ has been
tested with \Python\ 2.7 and \numpy\ 1.6
but it is likely to work with earlier versions of both of these as well.

An alternative installation method is to download the source code from
\url{http://dan.iel.fm/emcee} and run
\begin{lstlisting}
% python setup.py install
\end{lstlisting}
in the unzipped directory. Make sure that you have \numpy\ installed in this
case as well.

\section{Issues \& Contributions}

The development of \this\ is being coordinated on \github\ at
\url{http://github.com/dfm/emcee} and contributions are welcome. If you
encounter any problems with the code, please report them at
\url{http://github.com/dfm/emcee/issues} and consider
contributing a patch.

\section{Online Documentation}

To learn more about how to use \this\ in practice, it is best to check out the
documentation on the website \url{http://dan.iel.fm/emcee}. This page includes
the API documentation and many examples of possible work flows.

\end{document}